# Image-Guided Surgery: Technology, Quality, Innovation, and Opportunities for Medical Physics

J. H. Siewerdsen[1,2,3]

1. Departments of Imaging Physics, Radiation Physics, and Neurosurgery, The University of Texas MD Anderson Cancer Center
2. Institute for Data Science in Oncology, The University of Texas MD Anderson Cancer Center
3. Department of Biomedical Engineering, Johns Hopkins University

**Corresponding Author:**
Jeffrey H. Siewerdsen, PhD
Professor, Departments of Imaging Physics, Radiation Physics, and Neurosurgery
Co-Lead, Focus Area for Improved Safety, Quality, and Access to Cancer Care, Institute for Data Science in Oncology
The University of Texas MD Anderson Cancer Centner
Email: jhsiewerdsen@mdanderson.org

## ABSTRACT

The science and clinical practice of medical physics has been integral to the advancement of radiology and radiation therapy for over a century. In parallel, advances in surgery – including intraoperative imaging, registration, and other technologies within the expertise of medical physicists – have advanced primarily in connection to other disciplines, such as biomedical engineering and computer science, and via somewhat distinct translational paths. This review article briefly traces the parallel and convergent evolution of such scientific, engineering, and clinical domains with an eye to a potentially broader, more impactful role of medical physics in research and clinical practice of surgery. A review of image-guided surgery technologies is offered, including intraoperative imaging, tracking / navigation, image registration, visualization, and surgical robotics across a spectrum of surgical applications. Trends and drivers for research and innovation are traced, including federal funding and academic-industry partnership, and some of the major challenges to achieving major clinical impact are described. Opportunities for medical physicists to expand expertise and contribute to the advancement of surgery in the decade ahead are outlined, including research and innovation, data science approaches, improving efficiency through operations research and optimization, improving patient safety, and bringing rigorous quality assurance to technologies and processes in the circle of care for surgery. Challenges abound but appear tractable, including domain knowledge, professional qualifications, and the need for investment and clinical partnership.

## 1. PHYSICS AND ENGINEERING IN MEDICINE – A PARALLEL EVOLUTION

Advances in physics throughout the 20th century provided a foundation for the modern era of rapidly advancing technologies, computational capacity, materials, and other fields that are shaping the century ahead. The role of physics in 20th century medicine was particularly profound, with its genesis largely in the discovery and application of ionizing radiation for medical imaging and therapy. As a result, the evolution of radiology and radiation oncology to their modern forms proceeded hand-in-hand with medical physics, resulting in a natural, inseparable integration of medical physics research, education, and clinical practice within these disciplines. In the century ahead, the role of qualified medical physicists within the circle of care in radiology and radiation oncology is clear as agents of quality assurance (QA) and quality improvement (QI). Furthermore, successful translation of emerging technologies in radiology and radiation oncology – from medical physics or related quantitative sciences, such as biomedical engineering (BME), computer science, and data science – is also likely to proceed via medical physicists as agents of technology implementation, technical performance evaluation, and lifecycle management.

In contrast to radiology and radiation oncology, the evolution of modern surgery followed a path that is distinct, stemming from the barber and battlefield surgeons of the 18th and 19th centuries, approaching its modern form through development of antiseptic methods and anesthesia,[1] and spurred over the last 50 years in part by technologies arising from fields of engineering (e.g., biomedical, electrical, and mechanical) and materials science. As physicists like Roentgen and Curie are to radiology and radiotherapy, the engineers behind many of the major advances in modern surgery comprise a vast (and perhaps less immediately recognizable) playbill of engineering innovators, but their contributions have been no less profound in helping to shape the field. In the century ahead, these engineering fields are well positioned for breakthroughs in technology and data science that promise to transform surgery to a form that is more precise, quantitative, data-intensive, and quality-assured. A well-recognized challenge, however, is the arm's length at which the academic enterprises of engineering and computer science usually operate from clinical healthcare delivery systems – i.e., the hospital, and more specifically, the operating room (OR). Engineering innovation in the surgery domain is most often fueled by insight on unmet clinical needs gained by way of *formulated* (i.e., structured, usually transient) interactions between engineering researchers and trainees with surgeons,[2] providing engineers with invaluable perspective on clinical challenges, stakeholders, and practicalities. This approach to clinical insight and innovation stands in contrast to the *natural,* daily interaction of medical physicists in clinical practice with radiologists and radiation oncologists.

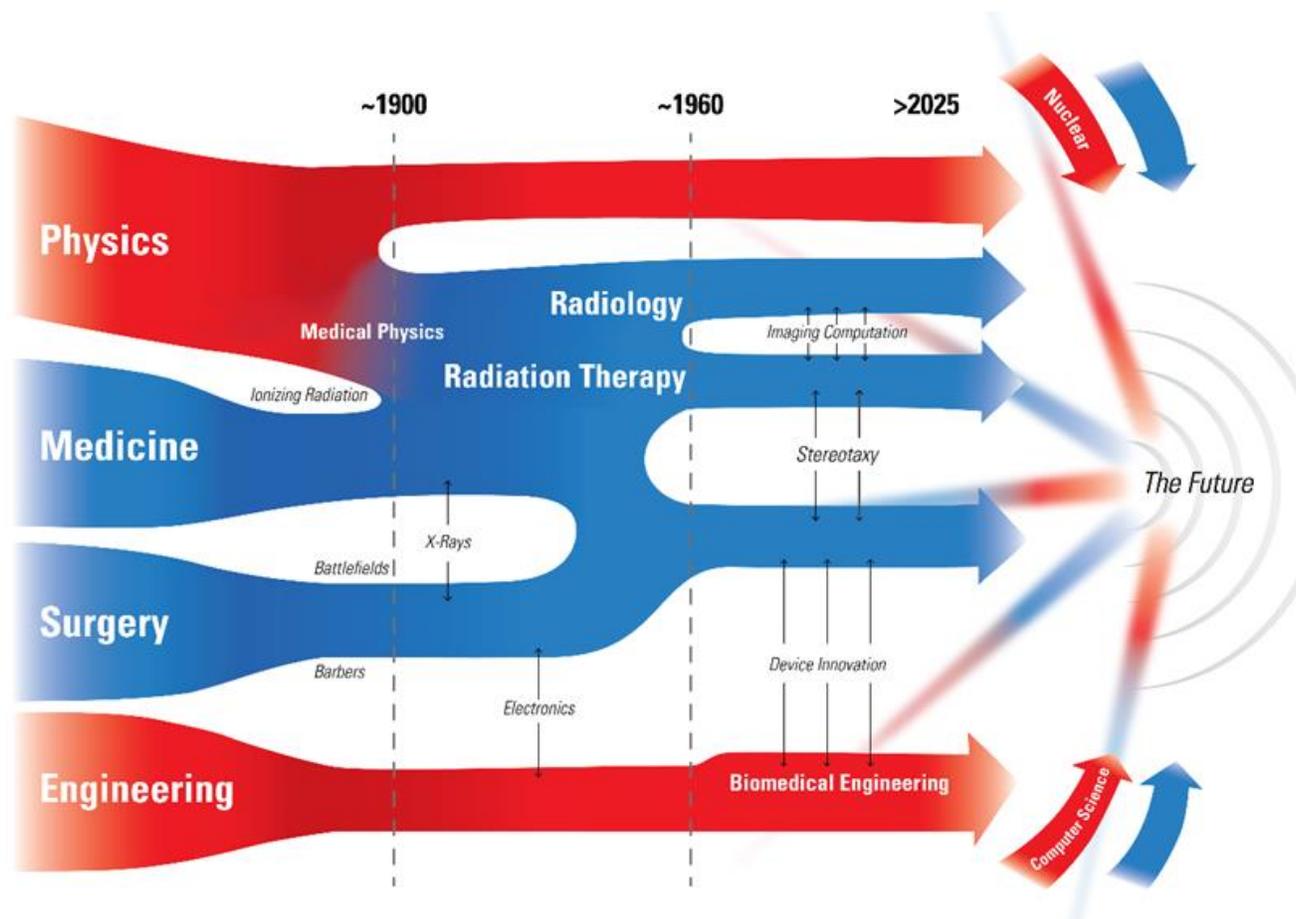

**Figure 1.** Qualitative depiction of evolution and interactions among physics, engineering, radiology, radiotherapy, and surgery. With the discovery of ionizing radiation at the end of the 19th century came the genesis of radiology and radiation oncology, with distinct specialization in the mid-1900s and medical physics integral to each. Medicine and surgery have long, distinct histories followed by convergence in culture, education, and professional practice in the 20th century. The emergence of BME as a recognizable domain after ~1960 presented an expansion of the engineering field (rather than a branching off as with medical physics). Numerous interactions among disciplines are noted, along with a future that may be both convergent and expansive with input from other disciplines, such as computer science. (Illustration provided by Vicky Soto, Creative Communications, The University of Texas MD Anderson Cancer Center.)

An exemplar of the contributions and relationship of engineering to clinical practice is Earl Bakken. Inspired as a young man by the technology of Dr. Frankenstein's laboratory, he was a garage innovator adept at maintaining and repairing electronic systems for medicine in Minneapolis MN. Through recognition of unmet clinical need combined with his own ingenuity and savvy for entrepreneurism and intellectual property, he started a small business in 1949 and developed and/or acquired systems for cardiac pacing, battery power, and other technologies that formed the foundation for what is now among the world's largest medical device companies (Medtronic, Minneapolis MN), with $32.4B revenue in 2024.[3] (By comparison to major companies in radiology and radiotherapy, 2024 revenue for Siemens Healthineers was $24.5B[4] and Varian was $4.1B.[5])

Meanwhile, over the last 60 years, the field of BME took shape in two distinct spheres – one dedicated to technology support in the hospital and one to academic research and education. The first is an essential workforce in daily operations of hospital technology and infrastructure not unlike Bakken's work with respect to the main technologies of interest (e.g., bedside monitors), observation of clinical challenges, and inventive problem solving. The second is an academic research and educational discipline with ~150 university departments in the USA,[6] granting >1200 doctorates per year in the USA,[7] but not commonly integrated with clinical practice. Rather, the academic enterprise of BME is marked by undergraduate and graduate education programs, extramurally funded research collaboration, academic-industry partnership, and startup for innovation and commercialization. Visiting the exhibit floor at two prominent conferences brings the contrast between fields to sharp focus. At the annual meeting of the Biomedical Engineering Society (BMES), hundreds of American universities exhibit in the market of academic recruitment: the education programs are the product, and the trainee is the customer. At the annual meeting of the American Association of Physicists in Medicine (AAPM), vendors exhibit the commercial market of medical imaging and radiation therapy: clinical systems are the product, and medical physicists (or radiologists, radiation oncologists, or hospital purchasing systems) are the customer.

Figure 1 attempts a qualitative depiction of this parallel evolution and interaction among fields. Physics and engineering share common origin in a more distant past – in the west, largely among thinkers in Babylonia and Greece. Fields of medicine and surgery similarly have more distant and distinct origins – via Hippocrates in medicine and unnamed practitioners attempting orthopaedic surgery and neurosurgery since paleolithic times.[1] Fast-forwarding to the 20th century in Fig. 1, note the major branch (and persistent degree of bifurcation from physics) for medical physics c. 1900 contributing to the genesis and evolution of radiology and radiation therapy, which in turn bifurcated by the mid-century to domains of diagnostic imaging and radiation oncology, respectively, but with medical physics integral to each. By contrast, the long history of surgery was fairly distinct from medicine prior to the 20th century before finding common cultural and educational foundation among physicians. Increasing levels of interaction between fields was evident over time, particularly between surgery and engineering. The mid-late 20th century saw the emergence of BME as a recognized domain in engineering – expanding (rather than bifurcating) the scope of engineering research and education. Looking ahead, one can anticipate many areas of convergence, expansion, and an ever-increasing dynamic of interactions among domains – including (rather than bifurcating) the scope of engineering research and education. Looking ahead, one can anticipate many areas of convergence, expansion, and an ever-increasing dynamic of interactions among domains – including "wet" sciences (molecular biology, genomics, etc.) and "dry" sciences (computer science, data science, etc.).

Given this backdrop of a century of parallel evolution in medical physics and engineering in medicine, this article first reviews some of the technologies that have emerged in support of image-guided surgery. Primary focus is on technologies for imaging and surgical guidance involving a geometric registration of image data to the patient and interventional device (as opposed to direct real-time visualization of anatomy and instrumentation as in video endoscopy). The effectors of surgical resection, device placement, or energy delivery are touched upon briefly in relation to image guidance, but are not the main focus, and the reader is referred elsewhere for more thorough review of technologies such as surgical robotics,[8] ablation,[9] and focused ultrasound.[10,11] We review the landscape of image-guided surgery applications, research funding, and educational paradigms, and conclude with an outlook and hypothesis concerning a potentially broader role of medical physicists within the circle of care of surgery as agents of quality, safety, and innovation.

## 2. IMAGE-GUIDED SURGERY TECHNOLOGIES

Numerous advances in medical imaging, registration, and real-time tracking over the last 25 years have enabled more minimally invasive, image-guided surgery in broad practice, with technologies developed and translated from industry and academic researchers working closely with clinical trailblazers. For purposes of definition herein, an "image-guided surgery" is one in which images acquired before or during the procedure are used to direct the placement of instruments and/or energy according to a geometric relationship between the images, surgical instruments, and the patient.

Two broad categories of image guidance can be appreciated – distinguished here in terms of how the image and world coordinates are co-localized. First is imaging in which both the anatomical structures of interest and the interventional device (or energy delivery) are directly visualized (together). The geometric relationship of the anatomy and instrument is thereby directly evident in the image, and "guidance" amounts to the surgeon's interpretation of the image and manipulation of the interventional device accordingly. To the extent that such scenarios involve use of an image – as opposed to visualization with the naked eye – they fall within our definition of "image-guided surgery" and include numerous examples – fluoroscopic guidance, ultrasound guidance, endoscopic video, and MR thermometry guidance – each depicting the anatomy and interventional device (or energy) directly in the same domain.

The second category is one in which the relationship of the image to the patient and interventional devices is inferred via geometric transformation rather than direct visualization. The image can be one acquired before the procedure (preoperative) or during the procedure (intraoperative), recognizing that the latter can provide updated visualization of tissue changes, deformation, the surgical resection, or implant. Preoperative imaging includes the full arsenal of diagnostic CT, MR, and nuclear medicine imaging. Intraoperative imaging can be accomplished by direct incorporation of such technologies in the OR – e.g., intraoperative CT[12] or MR[13] imaging – or modified forms of such technologies adapted to the intraoperative environment – e.g., increasing the diameter of the scanner bore, improving access, or faster scan protocols, as described below.

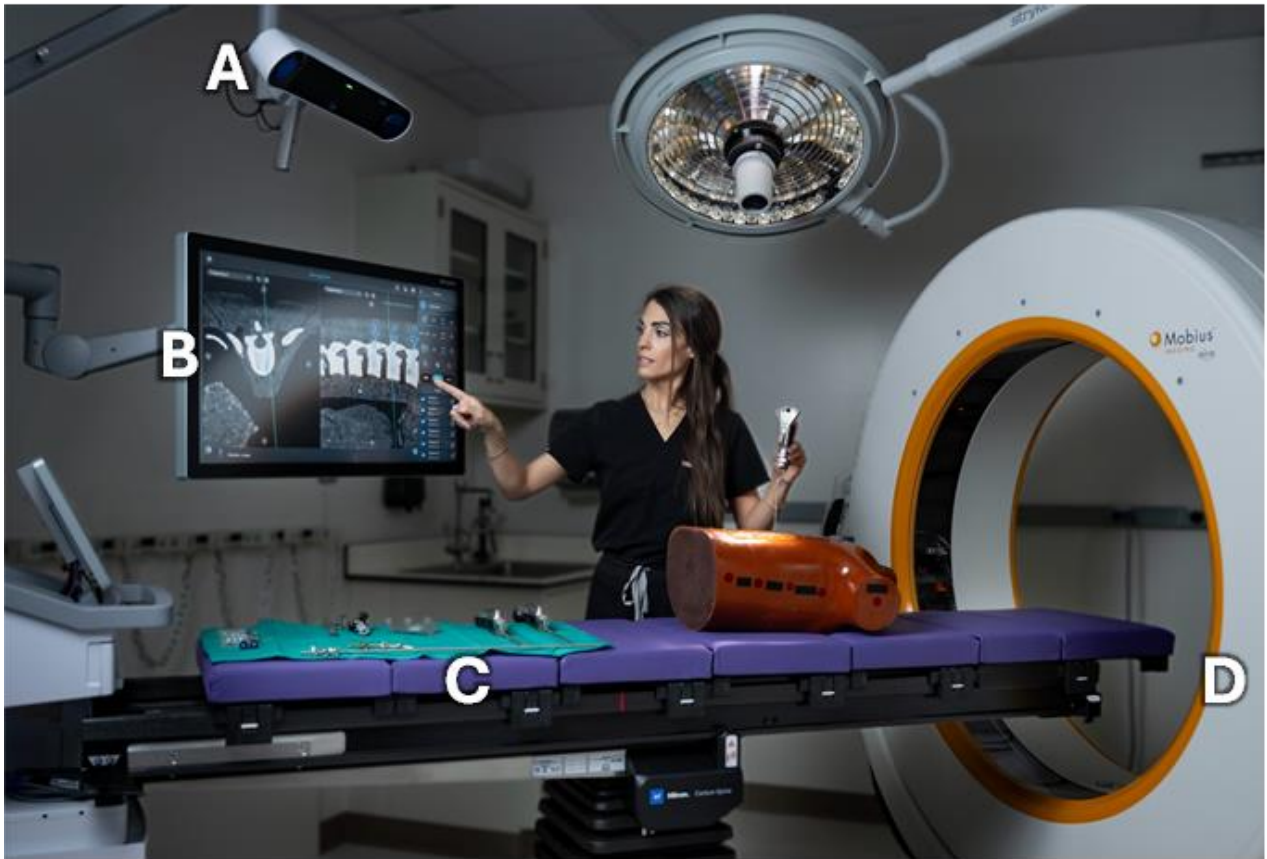

**Figure 2.** Photograph of surgical navigation system (Q Guidance; Stryker, Kalamazoo USA) in a laboratory environment: (A) stereoscopic optical (infrared) tracker; (B) navigation display, including image data (triplanar and volumetric views), planning information (e.g., spinal pedicle screw trajectories), and real-time visualization of tracked instrument positions; (C) surgical instrumentation, including tracked instruments with active or passive markers (e.g., the pointer, awl, gearshift, tap, and screwdriver); and (D) intraoperative imaging (mobile CT), also tracked by (A) via the 5 infrared markers visible across the top of the gantry. Photo courtesy of Mark Mulligan (The University of Texas MD Anderson Cancer Center) and Brandi Piercy (Stryker).

## 2.1 Surgical Tracking and Navigation

"Navigation" refers broadly to determining position and orientation according to a reference "map." In the image-guided surgery context, the map is the image [with coordinate frame $X_{image} = (x_{image}, y_{image}, z_{image})$], and positioning in the world is determined by real-time measurement with a tracker [with coordinate frame $X_{world} = (x_{world}, y_{world}, z_{world})$]. Surgical navigation relies on geometric registration of the image and world coordinate systems via an "image-to-world" transformation [14] relating $X_{image}$ to $X_{world}$, for example, by $X_{image} = T_{image}^{world} X_{world}$, where $T_{image}^{world}$ is a Euclidean transformation describing translation and rotation between the two coordinate frames. A simple means to determine $T_{image}^{world}$ uses co-localized "fiducial" marker points on the patient and visible in the image (e.g., cranial pins).[15] A surgical tracker is used to measure the position of a tracked instrument on 3 or more noncolinear points on the patient (recording the $X_{world}$ coordinates). By co-localizing the same points in the image ($X_{image}$ coordinates), the rigid transformation can be solved – for example, by minimizing RMSE distance between point pairs in the two coordinate frames. For intraoperative imaging systems (below), the image-to-world registration can also be solved by tracking the imaging system itself during the scan, since the location $X_{image}$ relative to the imaging system is known via calibration. Rigid displacements of the patient after registration can be accommodated by rigidly affixing a reference marker to the patient – e.g., to a head frame, a spinous process, or percutaneous pelvic pin.

Two categories of surgical tracker are commonly recognized: (1) optical / infrared; and (2) electromagnetic (EM). As illustrated in Fig. 2, the former includes passive (retroreflective) or active (emitting) markers placed in various arrangements on the surgical device and patient and sensed via a stereoscopic camera. Prevalent infrared trackers include those from Northern Digital Inc. (NDI, Waterloo Canada), such as the Polaris Vicra, Spectra, Lyra, and Vega systems, which have seen widespread implementation in research prototypes and commercial systems. Analogous clinical tracking systems include StealthStation (Medtronic, Minneapolis USA), Q Guidance (Stryker, Kalamazoo USA), and Curve (Brainlab, Munich Germany). EM trackers involve a field generator to produce a spatially varying EM field and small electric coils incorporated within a tracked device, with position inferred by the current induced in the coils. A common EM tracking system is the Aurora (NDI), with clinical embodiments including StealthStation EM (Medtronic), Scopis (Stryker), and Kick (Brainlab).

Such navigation assumes a stable, rigid relationship between the world and image coordinate frames. Even with a patient reference marker, rigid correspondence can degrade over the course of surgery due to unknown displacement of the reference marker relative to the patient or deformation of tissues relative to the image. Perturbation of the reference marker can be accommodated by updating the registration; however, nonrigid deformation of anatomy requires modeling of the deformation[16] and/or updating with an intraoperative image, as described further in the next section. Also described below, deformable image registration provides a means to maintain geometric correspondence of the surgical plan with respect to the patient – even in the presence of deformation – and has entered clinical use in image-guided radiation therapy;[17] however, registration in image-guided surgery still relies upon assumptions of rigid motion, and at the time of writing, surgical navigation systems operate under the assumption of a rigid relationship between the patient (reference marker) and the image.

## 2.2 Intraoperative Imaging

Imaging in the OR has been an important component of the surgical arsenal for well over a century, including the early use of x-ray imaging to localize surgical instruments relative to bone anatomy and foreign objects (e.g., shrapnel.[18] X-ray fluoroscopy on mobile C-arms is still a workhorse for orthopaedic and vascular surgery. Ultrasound imaging also offers valuable real-time imaging and device localization in a broad range of surgical interventions, including open or minimally invasive abdominal or pelvic surgery, orthopaedic surgery, and neurosurgery,[19] with uniform adoption limited in part by variations in surgeon training, preferences, and familiarity with ultrasound image visualization. By far the most prevalent real-time intraoperative imaging modality is

biomedical optical imaging, including a broad spectrum of endoscopy and fluorescence imaging systems.[20]

As mentioned above and illustrated in Fig. 3, adaptation of diagnostic CT and MR imaging to the OR presents major additional capabilities for 3D visualization of bone and soft tissues. Intraoperative CT systems include adaptations from each of the major diagnostic imaging concerns (Siemens, GE, Philips, and Canon) as well as mobile embodiments such as the Airo-CT (Stryker, Kalamazoo USA) and BodyTom (Samsung, Seoul Korea). A landmark of image-guided surgery research and clinical applications development is evident in the Advanced Multimodality Image Guided Operating Suite (AMIGO) operating theater at Brigham & Women's Hospital – a clinical research environment established in 2011.[21] The operating theater integrates imaging and guidance technologies across three adjoining rooms with in-line patient transport between CT, MRI, and the central OR with fluoroscopy, ultrasound, endoscopy, and navigation. Multidisciplinary research over the course of 20 years focused on technical aspects of image registration and visualization[22,23] as well as clinical applications ranging from orthopaedic surgery to brachytherapy.

The incorporation of flat-panel x-ray detectors in intraoperative fluoroscopy systems at the turn of the century[24] brought the capability for high quality cone-beam CT (CBCT) for image-guided interventions. Interestingly, the first medical applications of CBCT were in very different fields – dental imaging [24] and radiotherapy guidance[25] – and subsequent implementation on a mobile surgical C-arm[26] brought dual functionality of 2D fluoroscopy and 3D CBCT to the OR. Moreover, the familiar platform of a mobile C-arm carried relatively low cost consistent with applications in orthopaedic and spine surgery. A variety of mobile intraoperative CBCT systems have emerged over the last 20 years, including the O-arm (Medtronic, Minneapolis USA), Cios Spin (Siemens Healthineers, Forcheim Germany), Vision RFD (Ziehm, Nurembrerg Germany), and Loop-X (Brainlab, Munich Germany). With primary application in orthopaedic, spine, cranial, and vascular surgery, intraoperative CBCT remains a subject of interest for broader utilization in areas such as thoracic surgery,[27] head and neck surgery,[28] brachytherapy,[29] and image-guided transbronchial biopsy.[30] Integration with surgical tracking and navigation is fairly standard for each of these embodiments, allowing the navigation image to be updated during the course of surgery and reduce geometric errors imparted by nonrigid anatomical motion. In the interventional radiology context (somewhat outside the scope of this review), fixed-room angiographic C-arms regularly feature CBCT, and 3D imaging is broadly used in vascular interventions and image-guided ablation and embolization procedures.[31] Important areas of research in intraoperative CBCT include novel 3D image reconstruction methods,[32] 3D-3D and 3D-2D image registration,[33] and image acquisition from noncircular source-detector orbits.[34]

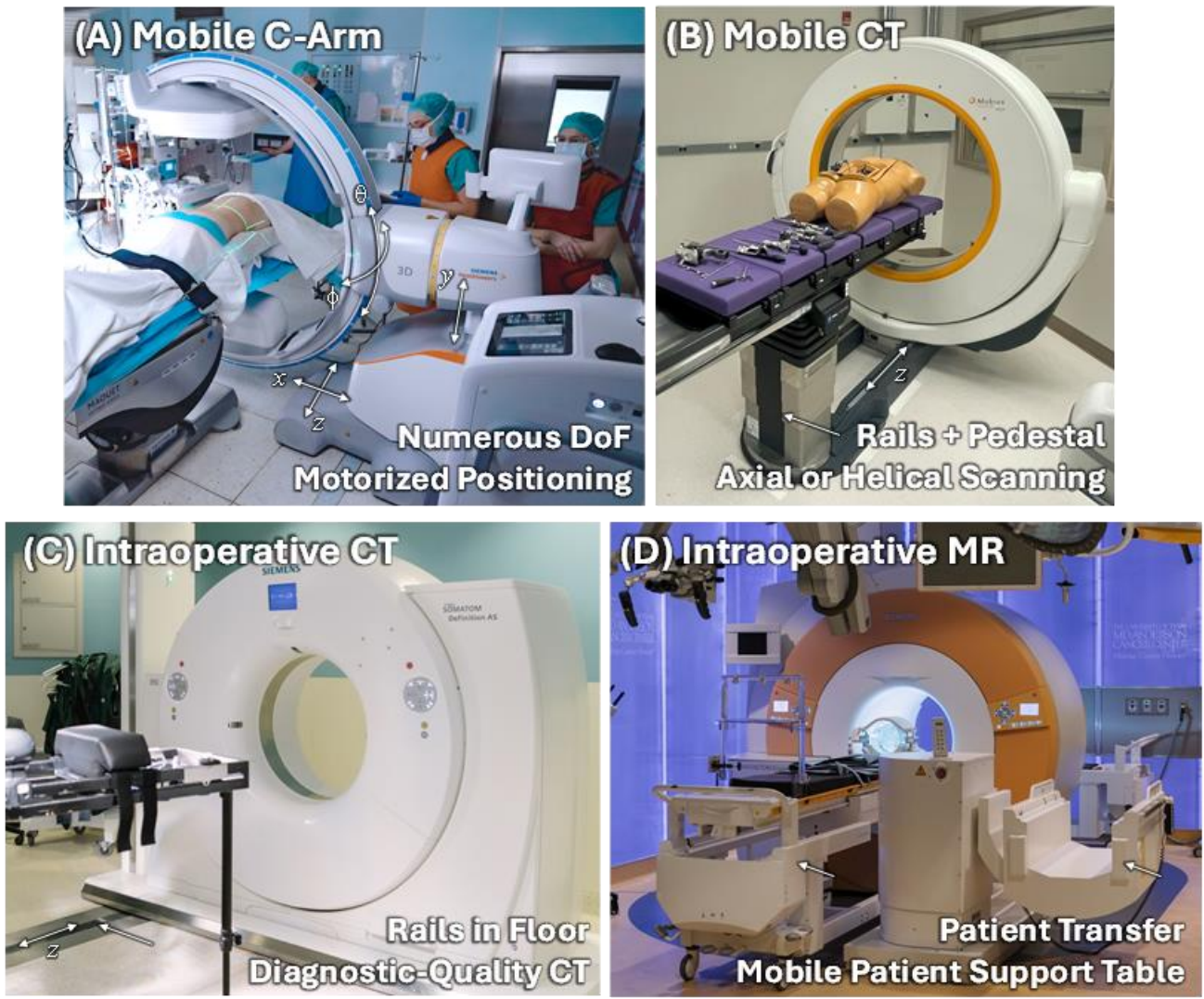

**Figure 3.** Intraoperative imaging systems, illustrating not only the spectrum of modalities but also the differences in mobile or fixed embodiments and implications for the patient support (operating table). (A) Mobile C-arm for 2D fluoroscopy and 3D CBCT (Ciartic Move, Siemens Healthineers). (B) Mobile intraoperative CT (Airo, Stryker) that acquires helical MDCT via translation of the gantry on rails that are physically integrated with a column and operating table. (C) Fixed-room intraoperative CT (Somatom Definition, Siemens Healthineers) that acquires helical MDCT via motion of the gantry on rails that are engineered within the floor. Intraoperative MR (Magnetom Espree, Siemens Healthineers) with requisite patient transfer and support structures. Photo credits: (A) Gerhard Kleinszig (Siemens Healthineers); (B-D) The University of Texas MD Anderson Cancer Center.

### 2.3 Image Registration

Intraoperative imaging permits the depiction of anatomy to be updated in a manner that reflects complex tissue motion and deformation, surgical resection, and device placement occurring during the procedure. As noted above, surgical navigation relies on an image-to-world registration combined with surgical tracking. In addition, bringing preoperative images, planning data, and/or previous intraoperative images into the most up-to-date geometric context requires some form of image-to-image registration.

Generally speaking, images can be registered according to a motion model, an objective function, and an optimization method by which the objective function is minimized (or maximized). A rigid motion model assumes 3 translational and 3 rotational degrees of freedom (+ scale for an affine model) between the moving and fixed images — appropriate, for example, to the cranium or a single bone. Piece-wise rigid motion models have also been reported for anatomical contexts such as the spine[35] or pelvis.[36] Nonrigid motion models expand the image registration capabilities to deformable tissues and have been reported in numerous surgical contexts, including the brain, head and neck, spine, lungs, liver, and other organs of the abdomen and pelvis.[37] Now commonly employed in image-guided radiation therapy,[38] deformable image registration is relatively nascent in image-guided surgery.

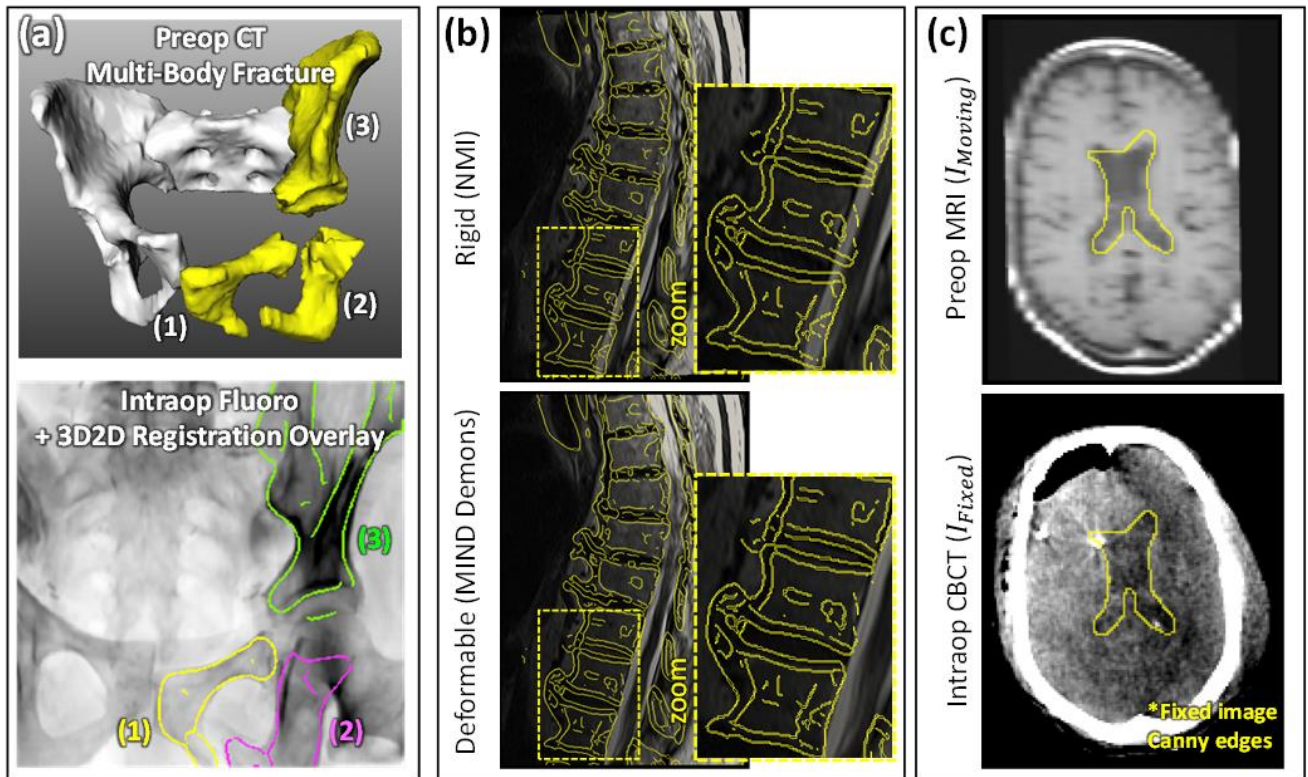

**Figure 3.** Examples of advanced image registration techniques in various image-guided surgery applications. (a) Multi-body 3D-2D registration of multi-body fractures in pelvic CT and fluoroscopy[39]. (b) 3D-3D multi-modality deformable registration of preoperative MRI and intraoperative CT via iterative optimization[40]. (c) 3D-3D multi-modality deformable registration of preoperative MRI and intraoperative CBCT via a deep learning approach for joint synthesis and registration (JSR)[41]. While the state-of-the-art in clinical use relies primarily on 3D-3D rigid image registration, approaches like these from various research contexts are steadily approaching translation to clinical use.

The objective function gives a measure of similarity between the moving (e.g., preop) and fixed (e.g., intraop) images, and maximizing the objective is intended to yield accurate geometric resolution between the images. For intra-modality registration (e.g., CT-to-CT), an intensity-based objective function such as mean-squared error (MSE) in voxel values may suffice. Inter-modality registration (e.g., MR-to-CT) warrant use of gradient-based or metrics such as gradient correlation (GC) or statistical metrics such as normalized mutual information (NMI). The optimization method is the means by which the objective function is minimized (or maximized), with classical methods such as gradient descent determining the direction at each step within an iterative process by which to align the images (in a rigid or local, nonrigid sense). Over the last decade, image registration using various forms of artificial neural network has emerged as a major area of research with applications throughout image-guided surgery, including deformable intra-modality and inter-modality registration.[42]

Examples in Fig. 3 illustrate some of the challenges and emerging solutions in image registration for image-guided surgery. Figure 3(a) illustrates a method for 3D-2D image registration (e.g., preoperative CT to intraoperative fluoroscopy) in which multiple rigid bodies (bone fragments) must be solved, whereas a conventional 3D-2D registration would involve a single, rigid transformation according to 6 degrees of freedom. The approach in Fig. 3(a) yields a solution in which the accuracy of reducing a multi-body fracture can be judged in fluoroscopy augmented by overlay of information registered from CT – e.g., Canny edge overlays of each registered bone fragment[39]. Figure 3(b) shows an example of 3D-3D deformable, multi-modality image registration in which preoperative MRI is registered to intraoperative CT for image-guided spine surgery. Deformation of the spine between preoperative and intraoperative imaging – e.g., due to variations in patient positioning or a purposeful, surgical modification of spinal alignment – may be accurate at the scale of a single vertebral level, but global alignment requires a non-rigid transformation. The NMI objective function handles the intensity mismatch between modalities, but a rigid solution leaves up to ~10 mm unresolved geometric error. A deformable solution such as the Demons algorithm with MIND (modality-insensitive neighborhood descriptor) similarity metric yields a deformable transform with ~2 mm accuracy[40]. Finally, a deep learning approach for deformable, multi-modality image registration is illustrated in Fig. 3(c), where preoperative MRI of the brain is to be accurately registered to intraoperative CBCT despite numerous challenges, including nonisotropic spatial resolution of the MRI, image quality limitations in CBCT, and strong, potentially non-diffeomorphic deformations of soft tissue (brain shift and loss of cerebrospinal fluid volume). The joint synthesis and registration (JSR) method illustrated in Fig. 3(c) uses MR-to-CT and CBCT-to-CT image synthesis to produce a common, intermediate, synthetic CT image domain via a generative adversarial network (GAN) in which the images are registered via a neural network decoder to estimate the deformation field[41]. At the time of writing, such solutions are largely in the research domain, and image registration in state-of-the-art image-guided surgery rely largely on rigid transformation.

## 2.4 Advanced Image Visualization

Image-guided surgery can quickly meet bottlenecks from information overload, raising the need for streamlined visualization, manipulation, and analysis of multi-modality 3D images and surgical planning data. Accomplishing such tasks in real-time in the operating room carries challenges that are distinct from 3D visualization in radiology reading or radiotherapy treatment planning, including hands-free operation and integration within an already crowded environment of towers, booms, equipment, and personnel about the patient. Extended reality (XR) systems offer the potential for advanced visualization that may be well suited to use in surgery.[43] For example, virtual reality (VR) systems are increasingly common in surgical education, permitting multiple operators to view and interact with 3D image information in a shared visual context and practice various tasks on digital models in complement to training on physical phantoms and cadavers.

Augmented reality (AR) holds even greater promise directly in the OR, where two distinct modes of AR visualization can be recognized. First is AR visualization that is unregistered to the world coordinate system – i.e., can be freely moved like a virtual boom. Such capability could improve ergonomics and situational awareness by bringing data within the surgeon's visual field, keeping their attention on the patient without straining to view information across multiple displays on booms or wall-mounted physical displays. Figure 4(A-B) shows such an example in thoracic surgery, where the surgeon uses AR (Apple Vision Pro) to view and manipulate multi-modality image, physiology, and endoscopy data within their field of view and without breaking scrub. A second category is AR visualization that is geometrically registered to the world coordinate system – similar to a tracked instrument, described above. In this context, image and planning data for surgical navigation can be viewed in a geometrically accurate manner directly on the surgical field, rather than on a separate display. Examples are shown in Fig. 4(C) for neurosurgery navigation and Fig. 4(D-E) for spine surgery navigation.

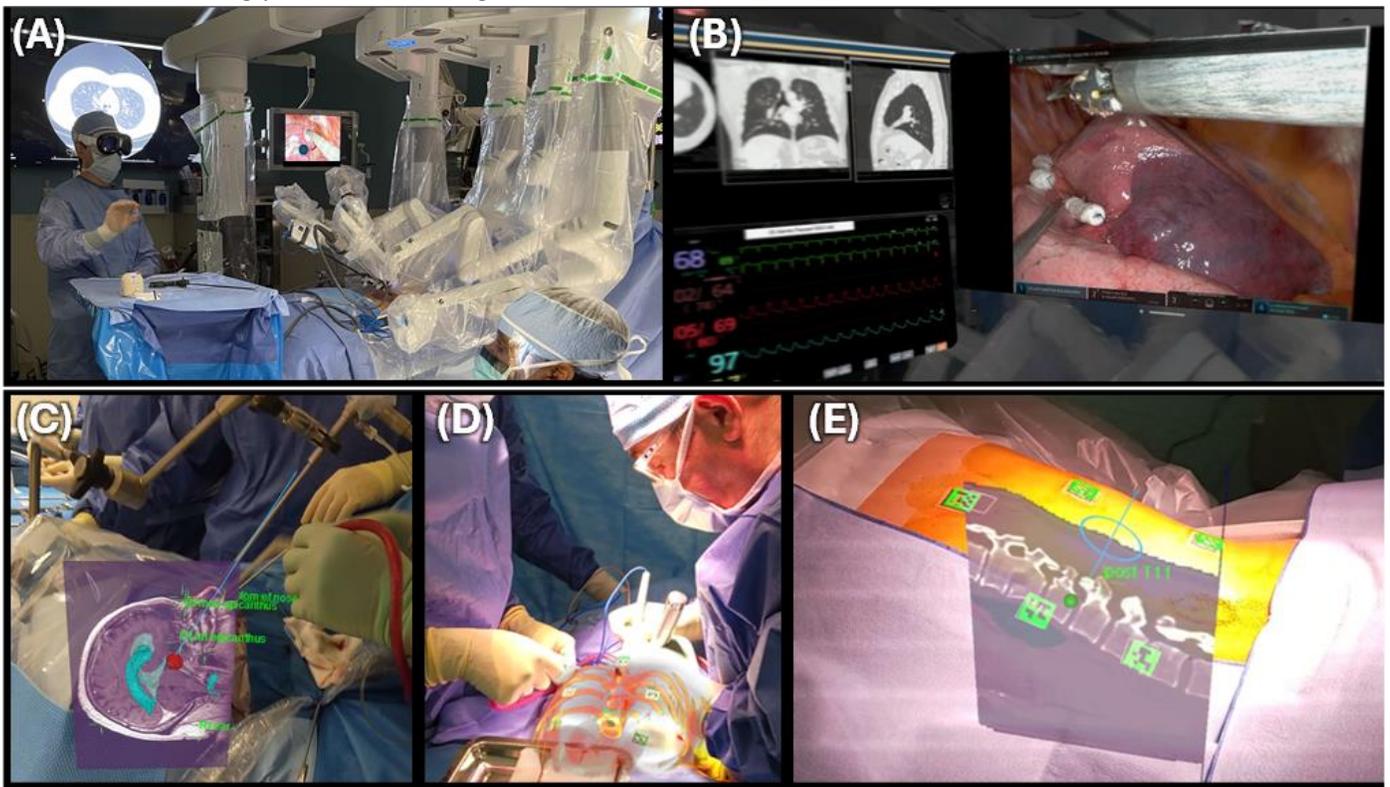

**Figure 4.** Augmented reality display in surgery. (A-B) A surgeon using augmented reality (Apple Vision Pro) in robot-assisted thoracic surgery for improved visualization of multi-modality data (tomography, physiology, and endoscopy) within his natural visual field (cf, multiple displays on booms). (C-D-E) Augmented reality (VisaRad, Novarad) display geometrically registered to the patient for visualization of image and planning data directly in the surgical field in (C) endoscopic skull base neurosurgery and (D-E) spinal neurosurgery. Photos (A-B) are courtesy of Dr. David Rice (The University of Texas MD Anderson Cancer Center) and (C-D-E) are courtesy of Dr. Gibby Wendell (Novarad).

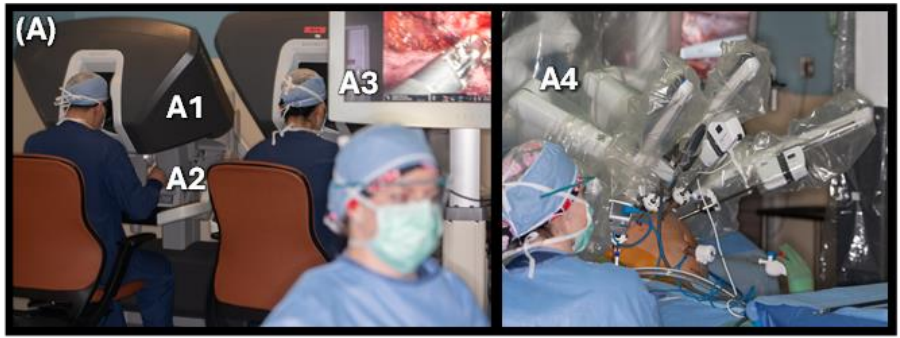
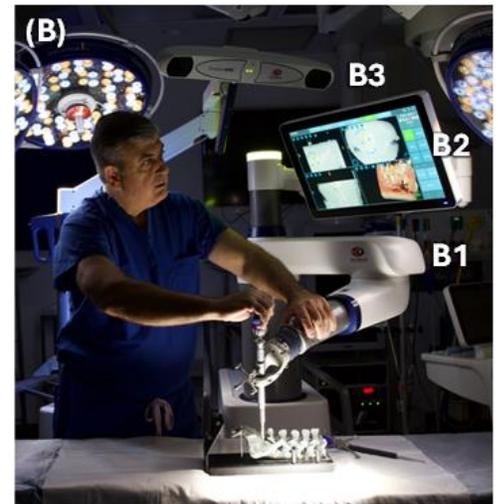

**Figure 5.** Example surgical robots. (A) Soft-tissue robot (DaVinci XI; Intuitive Surgical, Sunnyvale CA), including (A1) the console, (A2) dexterous fingertip manipulators, (A3) endoscopy display, and (A4) robotic arms for positioning and activation of intracorporeal / laparoscopic video and robotic end effectors. (B) Bone robot (Excelsius GPS; Globus Medical, Philadelphia PA), including (B1) robotic arm and mobile, floor-mounted console (cf, bed-mounted or patient-mounted) through which a surgical instrument is placed (e.g., pedicle screw driver), (B2) navigation display, and (B3) surgical tracker. Images in (A) are courtesy of Dr. Hop Tran Cao (The University of Texas MD Anderson Cancer Center), and image (B) is courtesy of Dr. Nicholas Theodore (Johns Hopkins University).

## 2.5 Robotic Assistance

As the imaging systems described above are an extension of the surgeon's eyes, so are robots an extension of his/her hands – and potentially tactile senses.[44] Although not a primary focus of this article, brief mention of this burgeoning technology is warranted, with two broad classes immediately recognized. The first are soft-tissue robots, with the DaVinci robot (Intuitive Surgical, Sunnyvale CA) illustrated in Fig. 5(A) as the most prominent example over the last two decades.[45] Such systems are typified by minimally invasive end effectors for soft tissue manipulation and resection complemented by endoscopic video (and sometimes ultrasound imaging) visualization. Arguably, such systems are not robots at all, operating instead via cooperative control of effectors via mechanical cables. Nor are such systems particularly "image-guided" as defined herein, relying instead on direct endoscopic visualization. Primary clinical applications are in surgeries of the abdomen and pelvis (e.g., prostatectomy, gynecological surgery, and liver surgery) with growing application in thoracic as well as head and neck surgery.

A second category is recognized in "bone" robots that operate extra-corporeally primarily as an assistant to orientation of surgical instruments – i.e., precise definition of a (linear) trajectory by which a device is to be implanted, such as a screw in the spinal pedicle or an electrode in the brain.[46] Figure 5(B) shows an example robotic assistant for spine surgery integrating with imaging and navigation. With some exceptions, such systems in current practice primarily operate as computer-controlled tool holders but do not directly place the implant, easing safety and regulatory considerations. Examples in spine and brain surgery include the Excelsius (Globus Medical, Audubon PA), Mazor-X (Medtronic, Minneapolis MN), Rosa (Zimmer, Warsaw IN), Mako (Stryker, Kalamazoo MI), and Cirq (Brainlab, Munich Germany). Such robots are invariably integrated with surgical tracking and navigation as described above, usually complemented by intraoperative imaging to account for intraoperative anatomical deformation.

## 2.6 Clinical Applications

The primary clinical application for which 3D imaging, registration, and navigation has transformed surgical capabilities over the last 25 years is intracranial neurosurgery, facilitating more minimally invasive approaches and guiding precise trajectories to surgical targets that avoid neurovasculature and minimize disruption of eloquent brain. Thanks to preoperative 3D imaging (MR and CT) and planning registered to intraoperative tracking, neuro-navigation permits neurosurgery to be performed in a more minimally invasive manner, and gone are the days in which a neurosurgeon approaches the region of interest but is unable to localize the target. Intraoperative imaging includes various fixed-room and mobile implementations of MR imaging (for direct visualization of the surgical target and surrounding tissues) as well as intraoperative CT and CBCT (typically in combination with rigid registration to preoperative images) as well as ultrasound imaging. Robotic assistance includes various "bone" robots of the types described above for holding a precise trajectory with respect to the patient (and image data) for placement of a biopsy needle, neuroelectrode, or other interventional device. The potential for nonrigid image registration to overcome deformation of the brain following incision of the dura is rapidly advancing – including surface-based[47] and volumetric[41] multi-modality image registration – as is the capability to fuse 3D image information with advanced visualization, extended reality systems,[48] and neuroendosocpy.[49]

Spine surgery has similarly seen increased adoption of 3D navigation, where a large market has helped to bear the cost of implementing advanced imaging, tracking, and robotics technologies in support of minimally invasive surgical techniques. Over the last 10 years, the field has witnessed increasing levels of integration from the conventional domain of spinal instrumentation among vendors to include well integrated systems for planning, navigation, intraoperative 3D imaging, and robotic assistance. Still, barriers to broad adoption persist, with 3D navigation advancing slowly but steadily beyond the ~20% market of early adopters. Hurdles associated with cost, complexity, and workflow are regularly cited, but increasing levels of automation, streamlined integration, and emerging technologies well integrated with clinical workflow – including 2D-3D image registration[50] and augmented reality[51] – complemented by evidence for improved safety and quality – will continue to reduce gaps in adoption.

It is natural to think that orthopaedic surgery presents a natural fit to 3D guidance, given the relatively rigid and high-contrast anatomy well suited to navigation via rigid registration and intraoperative fluoroscopy, CT, or CBCT. In the context of orthopaedic spine surgery, that is true. However, broad implementation and uptake has been somewhat challenged in areas such as joint replacement and trauma due to challenges with cost, complexity, and workflow, which are amplified orthopaedic contexts, where demands for rapid workflow are often high, and reimbursements can be relatively low. Still, major opportunity exists to improve safety, precision, more minimally invasive approaches, the quality of surgical product, and radiation exposure.

Applications development in other clinical areas presents a topic of ongoing interest, including thoracic surgery, abdominal surgery (e.g., liver and pancreas), and head and neck surgery, which would benefit from advances in intraoperative image quality and deformable image registration. Soft-tissue robotics (e.g., the DaVinci robot) combined with high-definition, stereo endoscopy have helped to transform such surgeries in the last 20 years and enable new surgical approaches. The benefits of intraoperative imaging in this context (e.g., using intraoperative CT or CBCT) will require steady advances in soft-tissue visualization, patient motion compensation, and deformable image registration. Meanwhile, one can anticipate an ongoing expansion of interventional radiology techniques in procedures that were once the exclusive domain of surgery via incorporation of navigation, new imaging technologies, and robotics in the interventional suite.

# 3. RESEARCH AND INNOVATION

Given the rapid pace and clinical impact of image-guided surgery technologies and applications reviewed above, it is unsurprising to see an increasing magnitude and scope of related research. For example, Fig. 6(a) shows a nonlinear increase in scientific publications on the topic of image-guided surgery over the last 25 years, presumably correlated with increasing NIH research funding summarized in Figs. 6(b-c). A strong uptick in ~2003 signaled a vibrant emergence of the field, with a 3x increase in NIH-funded projects between 2005 and 2024. Funding spikes (and standard deviation) about 2005 and 2010 are associated with large program awards. The NIH funding mechanisms supporting research and education in image-guided surgery over that time period (not shown for brevity) were primarily R01 research grants and R21 high-risk / high-reward grants, followed by R43 and R44 SBIR grants, evidencing the translational / commercializable nature of such work and showing the importance of industry partnership in technology research and development. Program grants are also prominent, as image-guided surgery research is multi-disciplinary by nature, often draws from disparate, major resources, and invites multi-institutional collaboration to bridge technical and clinical aims. Similar rationale makes the topic well suited to T32 training grants. NCI presented the largest source of funding over that time period, followed by NIBIB, recognizing that normalization by the size of each Institute's budget may tip in favor of NIBIB, NHLBI, or NINDS, consistent with the scope of clinical applications.

ARPA-H could add substantially to the scale, scope, and impact of such research funding. Compared to the ~300 awards and ~$350M disbursed over 4 years to image-guided surgery projects by NIH, the ARPA-H broad agency announcement in 2023 for Precision Surgical Interventions (PSI) offered awards to seven research groups[52] that could total up to $166.2M. Those projects aim to bridge anatomical and pathology scale imaging in the OR using technologies such as structured light microscopy, ultraviolet fluorescent microscopy, lightsheet microscopy, flexible microscope arrays, optical coherence tomography, photoacoustic imaging, fluorescence laparoscopy, and infrared and hyperspectral imaging.

At the time of writing, the outlook for trends in Fig. 6 uncertain. However, highly applied, translational, technology-oriented research that focuses physics and engineering efforts on clinically and commercially impactful challenges may be well positioned even in such an uncertain future. Many technology-oriented and data-intensive topics pertinent to image-guided surgery are potentially addressed within relatively short timelines, and they promise to reduce cost (e.g., by reducing costly adverse events), improve efficiency (e.g., through application of operations research principles), and generate commercial value (e.g., by translation of novel technologies) – themes that are more prevalent now than ever.

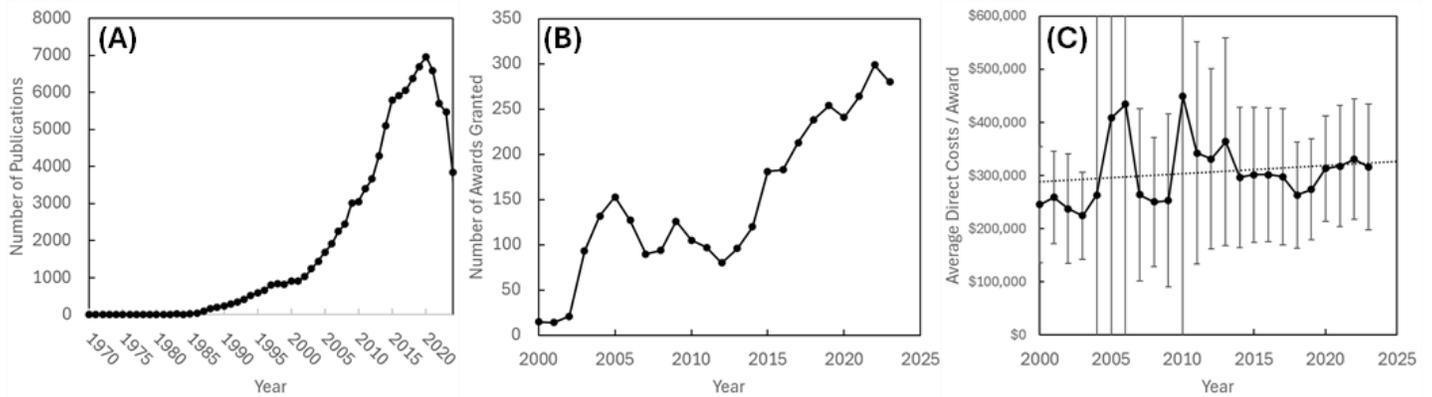

**Figure 6.** Increasing research activity in image-guided surgery in recent decades. (A) Number of scientific publications on the topic of image-guided surgery by year. (Source: PubMed; search term "image-guided surgery"). NIH funding of image-guided surgery research is summarized in terms of (B) number of awards granted (all types) and (C) average direct cost per award. (Source: NIH RePorter search keyword: image-guided surgery).

Academic-industry collaboration in image-guided surgery is another key ingredient to such research, combining the technical expertise of academic researchers, the practical insight of surgeons on key challenges, and the application-driven focus of industry scientists. Medical physicists can catalyze such activity by bridging the spectrum of technical expertise with the knowledge, capability, and access to clinical data and infrastructure to implement impactful solutions in practice.

A related dimension is the importance of entrepreneurship in bringing innovative technologies to positive impact in surgery. Unlike radiology or radiotherapy, where a small handful of established, multinational companies dominate the market, surgery presents a diverse, heterogeneous marketplace of commercial entities ranging from small startups with a single product to large corporations. Innovation in surgical technology often demonstrates early-stage proof-of-concept and feasibility in a relatively short, 1-2 year time scale, beyond which engineers learn to traverse the "valley of death" to commercial / clinical impact via startup, entrepreneur support, intellectual property, licensing, and navigation of regulatory pathways. This approach is evident in the training of biomedical engineers, where design, innovation, and entrepreneurship are often baked into ABET-approved curricula (e.g., bio-design programs[2]), compared to

CAMPEP-approved medical physics curricula, which focus primarily on thorough understanding of clinical practice rather than research and innovation.

## 4. COMPLEXITY, HETEROGENEITY, AND INTEROPERABILITY

The growing arsenal of intraoperative imaging, registration, navigation, visualization, and robotic technologies for image-guided surgery add to an already complex environment, including the surgical instrumentation, patient frames, operating table, anesthesia, and physiological monitoring. The result is an increasingly complex, heterogeneous, and variously well integrated conglomeration of technologies around the patient, presenting numerous challenges to daily workflow, which in turn inhibits further innovation.

***Complexity.*** Complexity in itself is not a problem; rather, it is the degree to which a complex system presents *complicatedness* to the user. The mobile phone offers an analogy – its complexity buried within a nearly featureless package – on the surface, uncomplicated. Similarly, image-guided surgery technologies involve complex inner workings, and major effort is warranted within and between technologies to streamline operation in a manner that buries distracting complicatedness and clearly surfaces aspects that improve the user's capabilities and performance. The point is not to overly simplify procedures that are inherently complex; rather, it is to bury unnecessary aspects of complicatedness.

***Heterogeneity.*** Neither is the heterogeneity of systems and vendors in itself a problem. Invoking the mobile phone analogy again, we see seamless functionality of countless apps from a broad range of vendors. Rather, it is the degree to which a heterogeneous system lacks *interoperability* among its components. In the surgery context: does the imaging system communicate with the navigation system, and with the robot, and with the surgical planning system, and each in turn with the PACS and EMR? Or are heroes called to the rescue for troubleshooting and workarounds at each connection point? Interoperability is key to managing the conglomeration and time-critical workflow of surgery, motivating communication standards[53] for surgery that are analogous to DICOM standards that revolutionized the interoperability of medical imaging systems, viewers, and data.

***Interoperability.*** As image-guided surgery technologies emerged in ~2005 – 2020, the need for streamlined workflow and interoperability became increasingly clear, as disparate specialized vendors advanced in the market. For example, image-to-world registration exercised increasingly automated methods rather than manual, paired-point registration that posed a major bottleneck to navigation workflow in early days – as an oft-repeated joke would have it, "slowly touch everything and lose two hours." Communication standards like FHIR and HL7 were another essential step toward integration and interoperability. More recently, the trend toward vertical integration among major market players is clear, evident first as strategic partnership among vendors, followed by internal development and, ultimately, total vertical integration of a single-vendor solution. The company who once only made the surgical implant now also produces the navigation system, the imaging system, the augmented reality display, and the robot.

***Reliance on Heroes.*** Along the way, a vital player emerged in the OR – the vendor reps – heroes at the ready to expertly assist and troubleshoot as issues arise. The overarching conglomeration is not their purview; rather, theirs is to ensure that their product functions properly. Consider, by contrast, the medical physicist in a radiation therapy department – cognizant and responsible for the safe, effective operation of systems throughout the treatment process – the imaging, treatment planning, dose calculations, and linear accelerator all within their purview.[54] Despite the complexity and heterogeneity of surgical technologies, there is currently no such actor in the OR, and the need for quality-assured operation of the system as a whole suggests an opportunity for medical physicists to expand and elevate their role in the surgical circle of care. Moreover, moving beyond such a strong reliance on heroes will require a more complete, systems view of the technologies and processes associated with each procedure.

## 5. OPPORTUNITIES FOR IMPACT

Despite the disparate evolution and influence of physics (in imaging and radiotherapy) and engineering (in surgery) traced in Fig. 1, one might optimistically anticipate convergence in image-guided surgery. After all, the distinction between "Imaging Physics" and "Therapy Physics" was naturally blurred as Radiation Oncology increasingly incorporated medical imaging for 3D planning and guidance, and medical physicists were integral to the emergence of image-guided radiotherapy (IGRT). Therefore, as Surgery increasingly incorporates medical imaging for planning and guidance, would not the role of medical physicists similarly apply? Despite the optimistic hypothesis, the answer may be "No." In contrast to the revolution of computerization, 3D imaging, and guidance in Radiation Oncology in the 1990s for which medical physicists were integral, the ongoing revolution in Surgery in the 2010s and 2020s has seen little such involvement. Rather, the technologies enabling image-guided surgery have grown primarily from engineering researchers and companies at arm's length from clinical practice, and their implementation and support has been provided by the vendors.

Technologies in the OR are one point of possible connection with medical physics; data is another. Looking ahead, the "OR of the Future" is one enabled not only by cutting-edge technologies but also the continuous capture, curation, and learning from the perioperative / intraoperative flow of data to improve safety, precision, efficiency, and complex decision making in surgery. Such capability is not just a desirable improvement, it may be a necessity. Just as the explosion of data science in Radiology and Radiation Oncology presented a challenge and opportunity for medical physicists to adapt, contribute to development, and knowledgably guide implementation, the disruptive change underway in Surgery presents an opportunity for medical physicists to engage.

Four broad areas in which medical physics expertise could be brought to bear in advancing safety, quality, and precision in Surgery are summarized below. Numerous challenges are also noted, including cultural and financial barriers along with gaps in domain expertise.

### 5.1 Technology Innovation

Medtech device development has been – and is likely to remain – a cornerstone of BME, including improved surgical devices, biomaterials, and signal processing. The talent pipeline of BME trainees opens perennially upon a vast spectrum of clinical challenges, including new and improved medical devices, wearable technologies, etc. Innovation in surgical robotics similarly leverages domain expertise founded in mechanical engineering, electrical engineering, and computer science, where foundational knowledge in kinematics, geometric transforms, and computer vision are de rigeur. Therefore, although surgical device innovation is not an area in which medical physicists regularly engage, it is familiar ground in engineering.

A broad space of technology innovation, however, would benefit from insights and expertise from medical physics, starting with those that involve delivery of energy to biological tissues. All modalities of intraoperative imaging present as a clear opportunity space, where imaging technologies, algorithms, analysis, and protocol optimization speak naturally to the domain expertise of medical physicists. Similarly, many areas of surgery require deformable registration of pre-, intra-, and post-operative images to gain the full benefit of such technologies. Therapeutics such as focused ultrasound, various forms of thermal ablation, and histotripsy are similarly well within reach of medical physics domain expertise. Furthermore, all of these technologies stand to benefit from rigorous implementation science approaches[55] to maximize clinical utilization and efficacy.

## 5.2 Surgical Data Science

Surgical data science has emerged in the last 5 years to help transform surgical intervention to a form that is increasingly data-driven, quantitative, and evidential in the quality of care.[56] Via measurement and modeling of surgical activities at multiple scales, from the local scale of the surgical incision to the macro scale of workflow about multiple ORs, it also aims to deliver more continuous capture, curation, and data-intensive learning. Surgical data science marks an important evolution in engineering and computer science approaches in relation to clinical applications – a connection for which arms-length innovation from academic departments outside the circle of care is likely insufficient, and deeper clinical connection holds tremendous value, including surgineering approaches for deepening the relationship of surgeons and quantitative scientists.[57,58]

The value of building a deep connection between quantitative scientists and clinical operations is well evidenced by a century of medical physics in Radiology and Radiation Oncology. Not surprisingly, these fields are now well positioned to leverage data science approaches in medical imaging and radiotherapy. Medical physicists have helped to establish the requisite data infrastructure and contributed the expertise and clinical experience that are essential to translating data science to clinical impact. Despite the gap between clinical domains, medical physicists may be well positioned to contribute to data science in surgery as well. Example areas for impact include infrastructure for improved data capture from the OR, standards for surgical data, structured reporting, outcomes measurement, image analysis (e.g., in surgical planning and intraoperative guidance), development of clinical decision support tools, and model deployment and life-cycle management.

## 5.3 Operations Research

Grounded in applied math and optimization, operations research is well established in areas of transportation, distribution, logistics, and other business sectors that have realized major gains in the efficiency and reliability of services over the last 2-3 decades. Application in healthcare, however, has been less, despite the potential to address major challenges, including schedule optimization, load forecasting, staffing, and other areas that could improve efficiency, reduce cost, raise patient and caregiver satisfaction, and improve access to healthcare. Many of the underlying methodologies are well established – e.g., schedule optimization[59] – and others could exercise novel machine learning methods in combination with large clinical datasets. Medical physicists are ideally positioned to access large datasets, contribute to development, and provide expert clinical perspective. Within the conventional domains of Radiology and Radiation Oncology, for example, challenges abound in which medical physicists could bring positive impact – for example: case load forecasting; robust schedule optimization in adaptive clinical workflows; concomitant requirements in staffing; and capacity planning for new facilities.

Outside these traditional domains, the opportunity for positive impact from operations research is similarly clear. In Medical Oncology, for example, medical physicists could exercise their expertise in mathematical optimization and data science to contribute to clinical decision support on medication plans that maximize patient-specific outcomes and treatment goals. Surgical scheduling presents another area for major improvement via optimization that could improve efficiency, OR utilization, wait times, and staffing.[60] The ability to derive such actionable insight presents an opportunity for medical physicists to elevate their contributions to highly reliable, quality care.

## 5.4 Quality and Safety

As agents of quality and safety, medical physicists lead the commissioning and periodic checks of imaging and treatment technologies and protocols in Radiology and Radiation Oncology that could result in major harms without such expert oversight and QA. They also lead the charge for improved QA methods that are more rigorous, quantitative, reproducible, and well suited to the modern arsenal of imaging and radiotherapy. In the similarly high-risk environment of the OR, the influence of medical physicists is relatively minor. Recognizing that surgery is among the top sources of preventable harm to the patient, with 1 in 3 patients suffering an adverse event,[61] major initiatives over the last 25 years have sought to address these major shortfalls, but progress has been slow.[62]

Although a medical physicist's expertise could certainly extend to intraoperative imaging, navigation, and robotics for image-guided surgery, their involvement in the deployment and management of such systems is relatively minor – perhaps limited to commissioning and registration of x-ray imaging systems, with regular QA somewhat of an afterthought compared to systems under their purview in diagnostic radiology. As a result, QA in the OR falls primarily to checks performed by technologists at system startup and service engineers in preventative maintenance. The risk of quality shortfalls is compounded by the fact that many of these systems are mobile and intended to service multiple ORs and types of procedures in a dynamic environment. Adaptation of QA methods to intraoperative imaging presents an immediate example in which medical physicists could elevate quality standards that are distinct from diagnostic and radiotherapy counterparts. Other intraoperative technologies similarly beckon as common points of failure or uncertainty for which problems are often detected only upon failure – e.g., surgical planning, navigation,[63] fluorescence imaging,[64] endoscopy, and robotics.[13,65–69] Additional insight on root cause and prediction of failure could be gained via failure mode and effects analysis (FMEA) and statistical process control (SPC).[70,71]

## 5.5 Challenges

The opportunities for medical physicists to bring positive clinical impact in surgery are accompanied by numerous cultural, epistemic, professional, and financial challenges. Such a broad spectrum of challenges are products of the natural evolution described in Section 1. Any one of them may be sufficient to nullify the hypothesis that medical physicists have a valuable role to play within the circle of care in surgery. Given the challenges of complexity, conglomeration, and interoperability surveyed in the Section 4, however, some form of change is needed to improve quality, safety, and workflow in the OR, constrain costs, and sustain the pace of impactful innovation enjoyed in the last century.

***Culture.*** Among the cultural challenges is a basic question: is it physics? A narrow view of medical physics pertaining to the application of radiation in medicine did not survive the first half of the 20th century, and fully within the physicist's purview are areas broadly covering all forms of medical imaging, the delivery of and response to novel therapeutics, computational modeling and measurement, and the numerous processes surrounding these activities.[54] All of these areas of the medical physics domain have analogous forms in surgery: the imaging modalities are similar, perhaps with increased use of optical modalities; many treatment technologies are analogous to those in interventional radiology; the planning and guidance technologies have counterparts in radiation therapy; and the surrounding processes of clinical workflow and dataflow are similarly complex. The question is therefore not "is it physics?" These topics are already recognized in medical physics. Rather, the question is whether the practice of medical physics can cross the gap to a new domain.

***Domain Knowledge.*** Other differences abound, including the diseases of interest, the pertinent treatment technologies, the clinical stakeholders, surgical workflow, as well as the terminology, rules, behaviors / etiquette, relationships, and values. There is therefore an epistemic gap between the traditional domain expertise of medical physics and the expertise required for the Surgery domain. Many of these topics simply are not part of traditional medical physics training or clinical practice – except in some instances in which the domains intersect – e.g., radiosurgery and neurosurgery. Meaningful involvement of medical physicists in the surgical circle of care would require a major expansion of the medical physics knowledge base.

***Qualification.*** This epistemic gap in turn raises questions for the professional practice, conventionally codified in terms of qualifications and licensure in Diagnostic, Therapeutic, and Nuclear Medicine specialties. Would contribution to the circle of care in surgery require a new specialization to support surgical services in a qualified, compliant manner? The answer with respect to intraoperative imaging is no, and even for imaging systems that are wholly distinct from their counterparts in radiology and radiation oncology (e.g., the O-arm or Airo-CT), the qualification that comes with specialization in Diagnostic Medical Physics is recognized. It is fair to consider, therefore, if characterization, commissioning, and QA of a surgical navigation system might be similarly within the purview of a qualified medical physicist specializing in Therapeutic Medical Physics who routinely performs such activities on radiotherapy tracking systems (e.g., ExacTrac, Brainlab, or Identify, Varian). The same could be argued with respect to a surgical robot (e.g., Excelsius, Globus, or Mako, Styker) relative to robotic radiation delivery platforms (e.g., CyberKnife, Accuray) that are within the medical physics wheelhouse. Alternatively, career paths for medical physicists in industry may grow – an expansion of the traditional "vendor rep" role to provide deeper medical physics expertise in support of imaging and registration technologies somewhat beyond the scope of conventional surgical device support.

***Investment.*** Financial considerations abound: will Surgery pay for services provided by medical physicists? Willing partnership between Radiology, Radiation Oncology, and

Surgery departments is a clear requirement, recognizing that each is a cost-constrained environment under considerable time and financial pressure. With the transition from fee-for-service to quality-based reimbursement, quality becomes the "currency of the realm"[72] – in surgery and elsewhere – and we can begin to evidence the value of a medical physicist in the circle of care:

- Considering the cost of OR time ($1000 - $8000 / hour[73]), what value is returned to Surgery by detecting a fault in an intraoperative CT or MR scanner prior to failure, permitting overnight service that saves at least one full day of surgeries?
- Considering the (widely variable) cost of each surgical case (>$10,000 / case), what value is gained by robust schedule optimization that frees contiguous time at the end of the day in one OR, permitting addition of one case / day, amounting to ~200 additional cases / year?
- What value can be ascribed to assuring reliable systems integration in the OR (e.g., the PACS and the surgical navigation system) that saves 10 min / case , amounting to ~1 hr / day otherwise lost to scrambling for a solution? Even if 1 hr / day saved does not permit an additional case / day, can one ascribe value to improved scheduling and wellness within a workforce facing major stress and burnout.[74]
- Furthermore, considering that the average cost per injury in surgery from preventable harm is nearly $60,000, what value is gained from verification of surgical planning and targeting that avoids 1 wrong-level spine surgery event (estimated to occur once in every 3000 spine procedures[75]) not to mention the operational / administrative cost in litigation? Similarly, what value is realized by avoiding 1 retained surgical instrument event (also estimated to occur once in every 3000 procedures[76–78]) by virtue of improved image quality with the mobile radiography system?

Aggregation of one or two of these propositions is comparable to the mean salary of a medical physicist in the US (~$200,000 / year[73]) The high cost of surgical services combined with the high cost of failure therefore slants the value proposition in favor of incorporating an expert within the circle of care who is dedicated to safety, quality, and problem solving across the complex systems comprising the surgical arsenal.

## 6. CONCLUSIONS

In reviewing aspects of image-guided surgery with an eye to opportunities for medical physics, this article briefly traced the disparate, parallel evolution of physics in radiology and radiotherapy relative to engineering in surgery. In their respective spheres, each contributed to major advances that have helped to shape the modern standard of care: the arsenal of modern technologies, algorithms, and measurements in medical imaging and radiation therapy; and the burgeoning space of intraoperative imaging, planning, guidance, visualization, and robotic assistance in surgery. Despite the somewhat distinct disciplines from which they have emerged, such technologies share a common ground in underlying the quantitative sciences of physics, engineering, applied mathematics, and computer science. If we agree that the intersection of physics and engineering in surgery could present a vibrant space of interaction in decades ahead, then how to engage as individuals and as a profession? Interdisciplinary collaboration and vision are key – identifying partners at the level of innovative projects, growing these to impactful programs, and ultimately rising to improved, sustainable improvements in surgical practice.

*SWOT.* That century of evolution produced two phenotypes that are now distinct in the scientific-medical-professional ecosystem: physicists within the circle of care in radiology and radiotherapy for whom clinical service is the principal component of the profession (and research is complementary); and engineers in industry or academic departments for whom research, innovation, and education are the driving forces (and clinical service is secondary). Each has adopted a strong role in their respective domain, each tinged with a weakness: medical physicists with unique domain expertise, but often challenged by time, resources, and/or incentive to drive impactful research; and engineers driving innovative research programs, often challenged in clinical insight, translation, implementation, and impact. In light of these strengths (S) and weaknesses (W), there is opportunity (O) – for medical physicists to have a greater role in surgical safety, quality, and technology innovation / implementation, and similarly for engineers to gain deeper insight on clinical practice, challenges, and workflow (i.e., surgineering). With respect to a SWOT analysis, one threat (T) is simply doing more of the same – consigning each to conventional domains and perpetuating existing shortfalls. With respect to workforce, clinical roles have not conventionally been a major career path for engineering or computer science trainees, but it will be interesting to see if that pattern holds in generations ahead. A more serious threat is a vacuous approach in which the influence of medical physics is expanded to new frontiers naively and brings no value, or worse.

*Quality Improvement.* At the level of individual effort and projects that could help medical physicists gain traction, QI projects can be a good starting point. Well within the medical physics wheelhouse, for example, are QI projects involving radiation exposure in the operating room – e.g., educating surgeons and other stakeholders in the OR on exposure levels, biological risk, and best-practice methods for exposure reduction – accompanied by monitoring and measurement to evidence exposure reduction and implementation of procedures to sustain the performance improvement. Implementation science[55] also presents a means by which medical physicists can bring their expertise to evaluation and deployment of new technologies in surgery – e.g., due diligence in survey of new intraoperative imaging technologies, working closely with vendors to identify workflow gaps, helping to steward change management and introduction in the OR, measuring performance during pilot phase and clinical deployment, and quantifying the value to clinical care. Success in QI and technology implementation is likely to propagate and would help to evidence the value of medical physicists in improving the quality of surgery.

***Education and Innovation.*** Many medical physics training programs already foster collaboration beyond radiology and radiation therapy – ranging from surgery to pathology – an approach with far-reaching positive implications in expanding the medical physics domain knowledge and influence. On the innovation front, medical physicists are also ideal mentors / supervisors of undergraduate interns or capstone project teams in engineering and computer science, where there is enormous appetite to work on practical challenges in a dynamic setting such as the hospital with diverse stakeholders. Opportunities to engage the next generation of engineers abound, including internships offered via university departments as well as the AAPM, BMES, or AIMBE. Engaging trainees from these disciplines not only gives students invaluable exposure to clinical challenges but also enriches the mentor's experience in topics that are outside traditional medical physics domains. Success at the scale of one-year projects can extend to deeper efforts spanning multiple years and generations of students, build interdisciplinary teams, yield preliminary data in support of larger projects, and make interactions between surgery and medical physics a normal aspect of clinical innovation.

***New Frontiers for Medical Physics.*** Recently, the AAPM charged an Ad Hoc Advisory Committee on New Science (AHNS)[79] to "evaluate likely changes in the application of physics in medicine over the next 10-30 years and to bring specific recommendations to the AAPM Board for investments or initiatives that can position us for success as an association and field." Among the strategic focus areas identified by the group were ways to deepen medical physics reach into other areas of medicine, including surgery. Strategies to extend the reach of medical physicists beyond traditional domains, included: engage leadership in academic / professional societies of surgeons, including the Society of Surgical Chairs and the Society of Neurological Surgeon; work with NIBIB to develop funding opportunities for physics in surgery; and exercise existing relationships with interventional radiology at intersections with surgery – e.g., interventional neuroradiology and endovascular neurosurgery. Further recommendations included: form an AAPM educational group on intraoperative imaging; show successful examples and stimulate new partnerships in clinically impactful research between surgeons, physicists, and data scientists – e.g., presentations at AAPM Annual Meeting, Spring Clinical Meeting, and/or Summer School; demonstrate the advantages of increased independence from vendor representatives in rigorous QA and QI; and identify opportunities for expanded medical physics training and QA of optical technologies (e.g., endoscopes and fluorescence-guided surgery).

Innovative (non-CAMPEP) medical physics fellowship programs and strategic hiring of medical physics faculty from outside conventional training channels could also help to advance the vision for an expanded role beyond conventional domains of diagnostic imaging and radiation therapy – certainly in connection to interventional radiology and cardiology and potentially to areas of surgery that increasingly rely on the imaging and registration technologies summarized above. Digital pathology similarly represents a potentially valuable frontier for medical physics involvement, with long-overdue integration with PACS and opportunities for imaging technology QA, development of more quantitative, reproducible biomarkers derived from pathology images, and registration between digital pathology and pre- and intra-operative imaging.

Thirty years ago, AAPM Task Group 40 published its report,[54] on comprehensive QA for radiation oncology, defining the roles and responsibilities of medical physicists with respect to specification, calibration, acceptance testing, commissioning, QA, measurements, calculations, supervision of maintenance, and education related to imaging, treatment planning, and treatment technologies in radiation oncology. The report also acknowledged that QA of the highest standard is likely accompanied by increased operating cost. Imagine, however, the costs and risks that would have been incurred in the course of such major advances in radiation therapy without expert, qualified oversight of these activities. Moreover, imagine the complexity and susceptibility to failure of such systems without medical physics expertise in the circle of care. It is possible that those advances would not have been feasible, with innovation paralyzed by unmanaged conglomeration, heterogeneity, and complexity, dependent on vendor representatives to assure that their particular link in the chain does not break.

The picture that comes to mind is akin to the current state of the operating room, which is similarly in an ongoing era of major advances in technology – and soon to be realized advances in surgical data science. Meanwhile, societies and individual leaders in the field of surgery look to a future in which the surgical craft is elevated to a quantitative, reproducible, and quality-assured practice. Within that outlook, there is clear opportunity for medical physicists to work in partnership with surgeons to help realize such a future, advance innovation and quality, enable further innovation, improve safety, reduce cost, and improve outcomes.

## ACKNOWLEDGMENTS

Numerous individuals provided insight and valuable discussion, including Dr. David Jaffray (The University of Texas MD Anderson Cancer Center), Dr. Rock Mackie (University of Wisconsin – Madison), colleagues at Johns Hopkins University (including Dr. Henry Brem, Dr. Nick Theodore, Dr. Ziya Gokaslan, Dr. Greg Osgood, Dr. Jay Khanna, and Dr. Michael Miller), Dr. Jonathan Irish (University of Toronto), collaborators at Siemens Healthineers (including Dr. Rainer Graumann, Gerhard Kleinszig, Dr. Sebastian Vogt, Dr. Clemens Bulitta, Dr. Tina Ehtiati, and Dr. Gouthami Chintalapani), Dr. Patrick Helm (Medtronic), Dr. Erin Angel (GE Healthcare), Dr. James Dobbins (Duke University), and colleagues in Imaging Physics (including Dr. Kyle Jones, Dr. Moiz Ahmad, Dr. Kristy Brock, and Dr. John Hazle) and the Division of Surgery (including Dr. Dr. Justin Bird, Dr. Laurence Rhines, Dr. Jeff Weinberg, Dr. David Rice, Dr. Hop Tran Cao, and Dr. Stephen Swisher) at The University of Texas MD Anderson Cancer Center. The author also extends thanks to numerous medical physics and biomedical engineering colleagues for feedback on perspectives conveyed in this work, biased though they may be from years spent in roughly equal parts between each field.

## BIBLIOGRAPHY

1. Ira Rutkow. *Empire of the Scalpel*. Simon and Schuster, 2022
2. Augustin DA, Denend L, Wall J, Krummel T, Azagury DE. The Biodesign Model: Training Physician Innovators and Entrepreneurs. In: 2019:71-88. doi: https://doi.org/10.1007/978-3-030-18613-5_7
3. Conor Stewart. Revenue of Medtronic from 2006 to 2024. https://www.statista.com/statistics/241515/total-revenues-of-medtronic-since-2006/
4. Revenue history for Siemens Healthineers from 2017 to 2024. https://companiesmarketcap.com/siemens-healthineers/revenue/.
5. Siemens posts strong revenues for Q4, fiscal year 2024. https://www.auntminnie.com/clinical-news/article/15707639/siemens-healthineers-siemens-posts-strong-revenues-for-q4-fiscal-year-2024#:~:text=Revenues%20in%20Siemens'%20Diagnostic%20segment,from%20%243.82%20billion%20last%20year.
6. Best Biomedical Engineering Programs. https://www.usnews.com/best-graduate-schools/top-engineering-schools/biomedical-rankings.
7. Roy J. Engineering by the Numbers. *American Society for Engineering Education*. Published online 2018:1-40.
8. Gomes P. Surgical robotics: Reviewing the past, analysing the present, imagining the future. *Robot Comput Integr Manuf*. 2011;27(2):261-266. doi: https://doi.org/10.1016/j.rcim.2010.06.009
9. Knavel EM, Brace CL. Tumor Ablation: Common Modalities and General Practices. *Tech Vasc Interv Radiol*. 2013;16(4):192-200. doi: https://doi.org/10.1053/j.tvir.2013.08.002
10. Maloney E, Hwang JH. Emerging HIFU applications in cancer therapy. *International Journal of Hyperthermia*. 2015;31(3):302-309. doi: https://doi.org/10.3109/02656736.2014.969789
11. ter Haar G. HIFU Tissue Ablation: Concept and Devices. In: 2016:3-20. doi: https://doi.org/10.1007/978-3-319-22536-4_1
12. Lunsford DL. Intraoperative Imaging with a Therapeutic Computed Tomographic Scanner. *Neurosurgery*. Published online 1984:559-561.
13. Black PM, Moriarty T, Alexander E 3rd, Stieg P, Woodard EJ, Gleason PL, Martin CH, Kikinis R, Schwartz RB, Jolesz FA. Development and implementation of intraoperative magnetic resonance imaging and its neurosurgical applications. In: *Neurosurgery*. 1997:831-845.
14. Cleary K, Peters TM. Image-Guided Interventions: Technology Review and Clinical Applications. *Annu Rev Biomed Eng*. 2010;12(1):119-142. doi: https://doi.org/10.1146/annurev-bioeng-070909-105249
15. Arun KS, Huang TS, Blostein SD. Least-Squares Fitting of Two 3-D Point Sets. *IEEE Trans Pattern Anal Mach Intell*. 1987;PAMI-9(5):698-700. doi: https://doi.org/10.1109/TPAMI.1987.4767965
16. Sotiras A, Davatzikos C, Paragios N. Deformable Medical Image Registration: A Survey. *IEEE Trans Med*

*Imaging*. 2013;32(7):1153-1190. doi: https://doi.org/10.1109/TMI.2013.2265603

17. Kaus MR, Brock KK. Deformable Image Registration For Radiation Therapy Planning: Algorithms And Applications. In: 2007:1-28. doi: https://doi.org/10.1142/9789812770042_0001

18. Mould RF. The early history of X-ray diagnosis with emphasis on the contributions of physics 1895-1915. *Phys Med Biol*. 1995;40(11):1741-1787. doi: https://doi.org/10.1088/0031-9155/40/11/001

19. Rozycki GS. Surgeon-Performed Ultrasound. *Ann Surg*. 1998;228(1):16-28. doi: https://doi.org/10.1097/00000658-199807000-00004

20. de Boer E, Harlaar NJ, Taruttis A, et al. Optical innovations in surgery. *British Journal of Surgery*. 2015;102(2):e56-e72. doi: https://doi.org/10.1002/bjs.9713

21. Jolesz FA. History of Image-Guided Therapy at Brigham and Women's Hospital. In: *Intraoperative Imaging and Image-Guided Therapy*. Springer New York; 2014:25-45. doi: https://doi.org/10.1007/978-1-4614-7657-3_2

22. Amla Z, Khehra PS, Mathialagan A, Lugez E. Review of the Free Research Software for Computer-Assisted Interventions. *Journal of Imaging Informatics in Medicine*. 2024;37(1):386-401. doi: https://doi.org/10.1007/s10278-023-00912-y

23. Gering DT, Nabavi A, Kikinis R, et al. An Integrated Visualization System for Surgical Planning and Guidance Using Image Fusion and Interventional Imaging. In: 1999:809-819. doi: https://doi.org/10.1007/10704282_88

24. Antonuk LE, Jee K -W., El-Mohri Y, et al. Strategies to improve the signal and noise performance of active matrix, flat-panel imagers for diagnostic x-ray applications. *Med Phys*. 2000;27(2):289-306. doi: https://doi.org/10.1118/1.598831

25. Jaffray DA. A personal perspective on the development and future of cone-beam CT for image-guided radiotherapy. *Med Phys*. 2023;50(S1):54-57. doi: https://doi.org/10.1002/mp.16238

26. Siewerdsen JH, Moseley DJ, Burch S, et al. Volume CT with a flat-panel detector on a mobile, isocentric C-arm: Pre-clinical investigation in guidance of minimally invasive surgery. *Med Phys*. 2005;32(1):241-254. doi: https://doi.org/10.1118/1.1836331

27. Setser R, Chintalapani G, Bhadra K, Casal RF. Cone beam CT imaging for bronchoscopy: a technical review. *J Thorac Dis*. 2020;12(12):7416-7428. doi: https://doi.org/10.21037/jtd-20-2382

28. Daly MJ, Siewerdsen JH, Moseley DJ, Jaffray DA, Irish JC. Intraoperative cone-beam CT for guidance of head and neck surgery: Assessment of dose and image quality using a C-arm prototype. *Med Phys*. 2006;33(10):3767-3780. doi: https://doi.org/10.1118/1.2349687

29. Polo A, Salembier C, Venselaar J, Hoskin P. Review of intraoperative imaging and planning techniques in permanent seed prostate brachytherapy. *Radiotherapy and Oncology*. 2010;94(1):12-23. doi: https://doi.org/10.1016/j.radonc.2009.12.012

30. Casal RF, Sarkiss M, Jones AK, et al. Cone beam computed tomography-guided thin/ultrathin bronchoscopy for diagnosis of peripheral lung nodules: a prospective pilot study. *J Thorac Dis*. 2018;10(12):6950-6959. doi: https://doi.org/10.21037/jtd.2018.11.21

31. Brock KK, Chen SR, Sheth RA, Siewerdsen JH. Imaging in Interventional Radiology: 2043 and Beyond. Radiology. 2023;308(1). doi: https://doi.org/10.1148/radiol.230146

32. Szczykutowicz TP, Toia G V., Dhanantwari A, Nett B. A Review of Deep Learning CT Reconstruction: Concepts, Limitations, and Promise in Clinical Practice. *Curr Radiol Rep*. 2022;10(9):101-115. doi: https://doi.org/10.1007/s40134-022-00399-5

33. Chen J, Liu Y, Wei S, et al. A survey on deep learning in medical image registration: New technologies, uncertainty, evaluation metrics, and beyond. *Med Image Anal*. 2025;100:103385. doi: https://doi.org/10.1016/j.media.2024.103385

34. Hatamikia S, Biguri A, Herl G, et al. Source-detector trajectory optimization in cone-beam computed tomography: a comprehensive review on today's state-of-the-art. *Phys Med Biol*. 2022;67(16):16TR03. doi: https://doi.org/10.1088/1361-6560/ac8590

35. Penney GP. *Registration of Tomographic Images to X-Ray Projections for Use in Image Guided Interventions.* 2000.

36. Han R, Uneri A, Ketcha M, et al. Multi-body 3D–2D registration for image-guided reduction of pelvic dislocation in orthopaedic trauma surgery. *Phys Med Biol*. 2020;65(13):135009. doi: https://doi.org/10.1088/1361-6560/ab843c

37. Darzi F, Bocklitz T. A Review of Medical Image Registration for Different Modalities. *Bioengineering*. 2024;11(8):786. doi: https://doi.org/10.3390/bioengineering11080786

38. Nenoff L, Amstutz F, Murr M, et al. Review and recommendations on deformable image registration uncertainties for radiotherapy applications. *Phys Med Biol*. 2023;68(24):24TR01. doi: https://doi.org/10.1088/1361-6560/ad0d8a

39. Deng B, Yao Y, Dyke RM, Zhang J. A Survey of Non-Rigid 3D Registration. *Computer Graphics Forum*.

2022;41(2):559-589. doi: https://doi.org/10.1111/cgf.14502

40. Han R, Uneri A, Vijayan R, et al. Fracture reduction planning and guidance in orthopaedic trauma surgery via multi-body image registration. *Med Image Anal*. 2021;68:101917. doi: https://doi.org/10.1016/j.media.2020.101917

41. Reaungamornrat S, De Silva T, Uneri A, et al. MIND Demons: Symmetric Diffeomorphic Deformable Registration of MR and CT for Image-Guided Spine Surgery. *IEEE Trans Med Imaging*. 2016;35(11):2413-2424. doi: https://doi.org/10.1109/TMI.2016.2576360

42. Han R, Jones CK, Lee J, et al. Joint synthesis and registration network for deformable MR-CBCT image registration for neurosurgical guidance. *Phys Med Biol*. 2022;67(12):125008. doi: https://doi.org/10.1088/1361-6560/ac72ef

43. Zhang J, Lu V, Khanduja V. The impact of extended reality on surgery: a scoping review. *Int Orthop*. 2023;47(3):611-621. doi: https://doi.org/10.1007/s00264-022-05663-z

44. Mayor N, Coppola AS, Challacombe B. Past, present and future of surgical robotics. *Trends in Urology & Men's Health*. 2022;13(1):7-10. doi: https://doi.org/10.1002/tre.834

45. Freschi C, Ferrari V, Melfi F, Ferrari M, Mosca F, Cuschieri A. Technical review of the da Vinci surgical telemanipulator. *The International Journal of Medical Robotics and Computer Assisted Surgery*. 2013;9(4):396-406. doi: https://doi.org/10.1002/rcs.1468

46. Yang HY, Seon JK. The landscape of surgical robotics in orthopedics surgery. *Biomed Eng Lett*. 2023;13(4):537-542. doi: https://doi.org/10.1007/s13534-023-00321-8

47. Frisken S, Unadkat P, Yang X, Miga MI, Golby AJ. Intra-operative Measurement of Brain Deformation. In: 2019:303-319. doi: https://doi.org/10.1007/978-3-030-04996-6_12

48. Dadario NB, Quinoa T, Khatri D, Boockvar J, Langer D, D'Amico RS. Examining the benefits of extended reality in neurosurgery: A systematic review. *Journal of Clinical Neuroscience*. 2021;94:41-53. doi: https://doi.org/10.1016/j.jocn.2021.09.037

49. Vagdargi P, Uneri A, Zhang X, et al. Real-Time 3-D Video Reconstruction for Guidance of Transventricular Neurosurgery. *IEEE Trans Med Robot Bionics*. 2023;5(3):669-682. doi: https://doi.org/10.1109/TMRB.2023.329245.

50. Zhang X, Uneri A, Huang Y, et al. Deformable 3D–2D image registration and analysis of global spinal alignment in long-length intraoperative spine imaging.

*Med Phys*. 2022;49(9):5715-5727. doi: https://doi.org/10.1002/mp.15819

51. Zhang J, Yang Z, Jiang S, Zhou Z. A spatial registration method based on 2D–3D registration for an augmented reality spinal surgery navigation system. *The International Journal of Medical Robotics and Computer Assisted Surgery*. 2024;20(1). doi: https://doi.org/10.1002/rcs.2612

52. ARPA-H announces awards to develop novel technologies for precise tumor removal . https://arpa-h.gov/news-and-events/arpa-h-announces-awards-develop-novel-technologies-precise-tumor-removal

53. Lemke HU, Vannier MW. The operating room and the need for an IT infrastructure and standards. *Int J Comput Assist Radiol Surg*. 2006;1(3):117-121. doi: https://doi.org/10.1007/s11548-006-0051-7

54. Kutcher GJ, LCMGWFHSLRJMJRP et al. Comprehensive QA for radiation oncology: report of AAPM radiation therapy committee task group . Published online 1994:581-581.

55. Krupinski EA. Translating computational innovations into reality: focus on the users! In: Tomaszewski JE, Ward AD, eds. *Medical Imaging 2023: Digital and Computational Pathology*. SPIE; 2023:507. doi: https://doi.org/10.1117/12.2663067

56. Maier-Hein L, Eisenmann M, Sarikaya D, et al. Surgical data science – from concepts toward clinical translation. *Med Image Anal*. 2022;76:102306. doi: https://doi.org/10.1016/j.media.2021.102306

57. Feussner H, Wilhelm D, Navab N, Knoll A, Lüth T. Surgineering: a new type of collaboration among surgeons and engineers. *Int J Comput Assist Radiol Surg*. 2019;14(2):187-190. doi: https://doi.org/10.1007/s11548-018-1893-5

58. Siewerdsen JH, Adrales GL, Anderson WS, et al. Surgineering: curriculum concept for experiential learning in upper-level biomedical engineering. *Int J Comput Assist Radiol Surg*. 2020;15(1):1-14. doi: https://doi.org/10.1007/s11548-019-02094-x

59. Cebi C, Atac E, Sahingoz OK. Job Shop Scheduling Problem and Solution Algorithms: A Review. In: *2020 11th International Conference on Computing, Communication and Networking Technologies (ICCCNT)*. IEEE; 2020:1-7. doi: https://doi.org/10.1109/ICCCNT49239.2020.9225581

60. Zhu S, Fan W, Yang S, Pei J, Pardalos PM. Operating room planning and surgical case scheduling: a review of literature. *J Comb Optim*. 2019;37(3):757-805. doi: https://doi.org/10.1007/s10878-018-0322-6

61. Adverse events affect over 1 in 3 surgery patients, US study finds. https://bmjgroup.com/adverse-events-affect-over-1-in-3-surgery-patients-us-study-finds/.

62. Sevdalis N, Hull L, Birnbach DJ. Improving patient safety in the operating theatre and perioperative care: obstacles, interventions, and priorities for accelerating progress. *Br J Anaesth*. 2012;109:i3-i16. doi: https://doi.org/10.1093/bja/aes391

63. Butz I, Fernandez M, Uneri A, Theodore N, Anderson WS, Siewerdsen JH. Performance assessment of surgical tracking systems based on statistical process control and longitudinal QA. *Computer Assisted Surgery*. 2023;28(1). doi: https://doi.org/10.1080/24699322.2023.2275522

64. Pogue BW, Zhu TC, Ntziachristos V, et al. AAPM Task Group Report 311: Guidance for performance evaluation of fluorescence-guided surgery systems. *Med Phys*. 2024;51(2):740-771. doi: https://doi.org/10.1002/mp.16849

65. Dieterich S, Cavedon C, Chuang CF, et al. Report of AAPM TG 135: Quality assurance for robotic radiosurgery. *Med Phys*. 2011;38(6Part1):2914-2936. doi: https://doi.org/10.1118/1.3579139

66. Cheon W, Cho J, Ahn SH, Han Y, Choi DH. High-precision quality assurance of robotic couches with six degrees of freedom. *Physica Medica*. 2018;49:28-33. doi: https://doi.org/10.1016/j.ejmp.2018.04.008

67. Hollander JB, Ibrahim IA, Petzel K. QUALITY ASSURANCE IN ROBOTIC PROSTATECTOMY AT A MULTI-USER COMMUNITY HOSPITAL. *Journal of Urology*. 2008;179(4S):607-607. doi:10.1016/S0022-5347(08)61778-4

68. Georgeson K. Surgical innovation and quality assurance: Can we have both? *Semin Pediatr Surg*. 2015;24(3):112-114. doi: https://doi.org/10.1053/j.sempedsurg.2015.02.007

69. Siampli E, Monfaredi R, Pieper S, Li P, Beskin V, Cleary K. A standardized method for accuracy study of MRI-compatible robots: case study- a body-mounted robot. In: Fei B, Linte CA, eds. *Medical Imaging 2020: Image-Guided Procedures, Robotic Interventions, and Modeling*. SPIE; 2020:95. doi: https://doi.org/10.1117/12.2550575

70. Thornton E, Brook OR, Mendiratta-Lala M, Hallett DT, Kruskal JB. Application of Failure Mode and Effect Analysis in a Radiology Department. *RadioGraphics*. 2011;31(1):281-293. doi: https://doi.org/10.1148/rg.311105018

71. Giardina M, Cantone MC, Tomarchio E, Veronese I. A Review of Healthcare Failure Mode and Effects Analysis (HFMEA) in Radiotherapy. *Health Phys*. 2016;111(4):317-326. doi: https://doi.org/10.1097/HP.0000000000000536

72. Morrow R. *Leading High-Reliability Organizations in Healthcare*.; 2016.

73. Childers CP, Maggard-Gibbons M. Understanding Costs of Care in the Operating Room. *JAMA Surg*. 2018;153(4):e176233. doi: https://doi.org/10.1001/jamasurg.2017.6233

74. Nagle E, Griskevica I, Rajevska O, Ivanovs A, Mihailova S, Skruzkalne I. Factors affecting healthcare workers burnout and their conceptual models: scoping review. *BMC Psychol*. 2024;12(1):637. doi: https://doi.org/10.1186/s40359-024-02130-9

75. Mody MG, Nourbakhsh A, Stahl DL, Gibbs M, Alfawareh M, Garges KJ. The Prevalence of Wrong Level Surgery Among Spine Surgeons. *Spine (Phila Pa 1976)*. 2008;33(2):194-198. doi: https://doi.org/10.1097/BRS.0b013e31816043d1

76. Bani-Hani KE, Gharaibeh KA, Yagha RJ. Retained Surgical Sponges (Gossypiboma). *Asian J Surg*. 2005;28(2):109-115. doi: https://doi.org/10.1016/S1015-9584(09)60273-6

77. Zahiri HR, Stromberg J, Skupsky H, et al. Prevention of 3 "Never Events" in the Operating Room: Fires, Gossypiboma, and Wrong-Site Surgery. *Surg Innov*. 2011;18(1):55-60. doi: https://doi.org/10.1177/1553350610389196

78. Egorova NN, Moskowitz A, Gelijns A, et al. Managing the Prevention of Retained Surgical Instruments. *Ann Surg*. 2008;247(1):13-18. doi: https://doi.org/10.1097/SLA.0b013e3180f633be

79. Samei E, Pawlicki T, Bourland D, et al. Redefining and reinvigorating the role of physics in clinical medicine: A Report from the <scp>AAPM</scp> Medical Physics 3.0 Ad Hoc Committee. *Med Phys*. 2018;45(9). doi: https://doi.org/10.1002/mp.13087